\documentclass[%
reprint,
 amsmath,amssymb,
pra,
floatfix,
]{revtex4-2}

\usepackage{graphicx}
\usepackage{dcolumn}
\usepackage{bm}
\usepackage{xcolor}
\usepackage{hyperref}
\usepackage{layouts}
\usepackage{multirow}

\usepackage{pifont}
\usepackage{siunitx}

\begin{document}

\preprint{APS/123-QED}

\title{Geometry dependence of TLS noise and loss in \textit{a}-$\mathrm{SiC}$:$\mathrm{H}$ parallel plate capacitors for superconducting microwave resonators}

\author{K. Kouwenhoven$^{1,2}$}
\email{k.kouwenhoven@sron.nl}
\homepage{terahertz.tudelft.nl}

\author{G.P.J. van Doorn$^{1}$}
\author{B.T. Buijtendorp$^{2}$} 
\author{S.A.H. de Rooij$^{1,2}$}
\author{D. Lamers$^{1}$}
\author{D.J. Thoen$^{1}$}
\author{V. Murugesan$^{1}$}
\author{J.J.A. Baselmans$^{1,2}$}
\author{P.J. de Visser$^{1}$}

\affiliation{%
 $^1$Netherlands Institute for Space Research (SRON), Niels Bohrweg 4, Leiden 2333 CA, Netherlands
 }%
\affiliation{%
 $^2$Department of Microelectronics, Delft University of Technology, Mekelweg 4 2628 CD, Delft, Netherlands
 }%

\date{\today}


\begin{abstract}
Parallel plate capacitors (PPC) significantly reduce the size of superconducting microwave resonators, reducing the pixel pitch for arrays of single photon energy-resolving kinetic inductance detectors (KIDs).
The frequency noise of KIDs is typically limited by tunneling Two-Level Systems (TLS), which originate from lattice defects in the dielectric materials required for PPCs.
How the frequency noise level depends on the PPC's dimensions has not been experimentally addressed.
We measure the frequency noise of 56 resonators with \textit{a}-$\mathrm{SiC}$:$\mathrm{H}$ PPCs, which cover a factor 44 in PPC area and a factor 4 in dielectric thickness.
To support the noise analysis, we measure the resonators' TLS-induced, power-dependent, intrinsic loss and temperature-dependent resonance frequency shift. 
From the TLS models, we expect a geometry-independent microwave loss and resonance frequency shift, set by the TLS properties of the dielectric.
However, we observe a thickness-dependent microwave loss and resonance frequency shift, explained by surface layers that limit the performance of PPC-based resonators.
For a uniform dielectric, the frequency noise level should scale directly inversely with the PPC area and thickness.
We observe that an increase in PPC size reduces the frequency noise, but the exact scaling is, in some cases, weaker than expected.
Finally, we derive engineering guidelines for the design of KIDs based on PPC-based resonators.
\end{abstract}

\maketitle

\section{Introduction}
\label{sec:Introduction}

Superconducting microwave resonators are one of the key elements of kinetic inductance detectors (KIDs) \cite{Day2003,Zmuidzinas2012,Baselmans2022} and  superconducting qubits \cite{Steffen2017,Krantz2019,Kjaergaard2020}.
The current resonators are typically based on planar structures such as coplanar waveguides (CPWs) and interdigitated capacitors (IDCs).
These planar structures provide little capacitance per unit area since the fields are spread between the substrate and air, which limits the packing density or pixel pitch of these resonators \cite{Meeker2018,Kouwenhoven2023}.
An alternative is the parallel-plate capacitor (PPC) \cite{Weber2011,Boussaha2019,Beldi2019,Zotova2023}, for which the entire field is in a dense ($\epsilon_r \sim 10$) and/or thin dielectric layer.
Such a PPC can drastically shrink the resonator's size, but the deposited dielectric required to fabricate a PPC will likely increase the resonator's microwave loss and frequency noise level due to tunneling states in the dielectric.

The microscopic nature of these two-level tunneling states (TLSs) is still unknown, but they are typically assumed to arise from disorder in the crystalline lattice.
Due to the disordered lattice, one or multiple atoms can tunnel between two energetically similar states modeled by the standard tunneling model (STM) \cite{Phillips1972,Gao2008_semiemperical,Gao2008thesis}.
The TLSs can couple to the electric field of the resonator through their electric dipole moment and modify the material's dielectric constant based on their individual states.
Averaging over all TLSs that are resonant with the resonator gives the TLS contribution to the dielectric constant, $\epsilon_{\rm{TLS}}$ \cite{Gao2008_semiemperical, Gao2008thesis}.

The real part of $\epsilon_{\rm{TLS}}$ introduces a resonance frequency shift, while the imaginary part introduces microwave loss. 
In addition, TLSs can randomly switch states, which gives rise to a time-fluctuating dielectric constant.
This, in turn, causes the resonator's resonance frequency to fluctuate in time, introducing excess frequency noise. 

Recent studies have focused on developing and characterizing low-loss and low-noise dielectrics \cite{OConnell2008, Buijtendorp2022}.
However, how the microwave loss and frequency noise level scale with a PPC's dimensions has not been experimentally addressed.

Here, we experimentally study the TLS noise of PPC resonators based on hydrogenated amorphous silicon carbide (\textit{a}-$\mathrm{SiC}$:$\mathrm{H}$).
Recent work showed that \textit{a}-$\mathrm{SiC}$:$\mathrm{H}$, depending on the exact deposition details, can be a low-loss, low-stress, deposited dielectric \cite{Buijtendorp2022}.
The exact deposition details for \textit{a}-$\mathrm{SiC}$:$\mathrm{H}$ used in this work are found in Ref.~\cite{Buijtendorp2022}.
We vary two geometrical parameters of the PPC: the plate area and the dielectric thickness.
We present an extensive dataset, containing 56 devices, with a factor of 44 in area variation and a factor of 4 in dielectric thickness.

In Sec.~\ref{sec:design} we describe the resonator design, fabrication and the experimental setup.
In Sec.~\ref{sec:S21} we focus on the TLS-induced microwave loss and frequency shift and show that the dielectric between the PPC plates contains lossy surface layers.
Sec.~\ref{sec:Noise} focuses on the resonator's frequency noise and how it scales with the PPC's dimensions.
We first discuss the frequency noise spectrum and the power and temperature dependencies.
Then, we show that the noise level decreases when the area or thickness of the PPC increases.
In Sec.~\ref{sec:discussion} we draw conclusions on the practical applicability of PPC-based resonators and discuss PPC-based KIDs.

\section{Design, Fabrication and experimental setup}
\label{sec:design}
\begin{figure}[hbt]
  \includegraphics{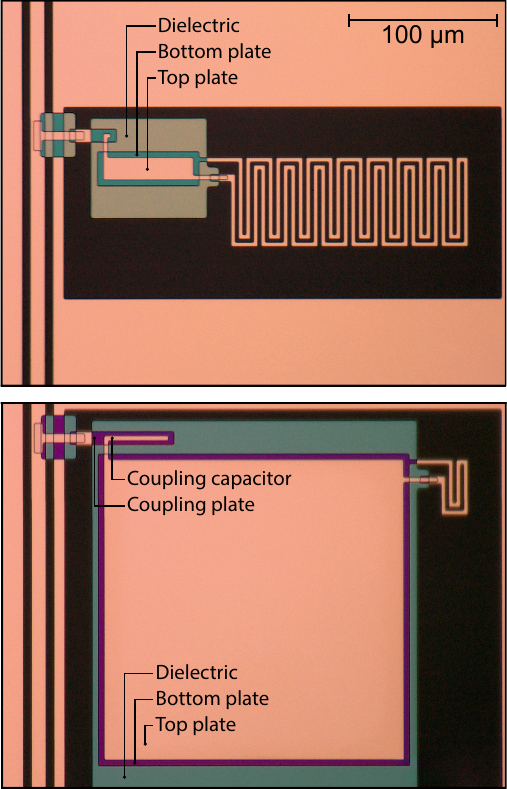}
  \caption{Optical micrograph of lumped element resonators with \textit{a}-$\mathrm{SiC}$:$\mathrm{H}$ parallel-plate capacitors. All metallic structures (pink) are $\mathrm{NbTiN}$ with a $T_{\rm{c}}$ of 14.3 K. Perceived colour of \textit{a}-$\mathrm{SiC}$:$\mathrm{H}$ is different if \textit{a}-$\mathrm{SiC}$:$\mathrm{H}$ is on top of the $\mathrm{NbTiN}$ layer (i.e. on top of the bottom plate) and varies due to thickness variations over the wafer. \textbf{Top}, smallest (60$\times$16 \SI{}{\micro\meter\squared}) PPC resonator; \textbf{bottom}, Biggest (200$\times$200 \SI{}{\micro\meter\squared}) PPC resonator.}
  \label{fig:PPC_KIDS}
 \end{figure}

We use a lumped-element resonator design consisting of an \textit{a}-$\mathrm{SiC}$:$\mathrm{H}$ PPC and a meandered inductor, as shown in Fig.~\ref{fig:PPC_KIDS}.
The resonators couple to the CPW readout line with a small separate capacitor with a bottom plate that is galvanically connected to the central line of the CPW.
The two ground planes of the CPW are electrically balanced at regular intervals by bridges that galvanically connect the ground planes (not shown).
The resonators, coupling lines, readout line, and bridges are made from $\mathrm{NbTiN}$ with a critical temperature ($T_{\rm{c}}$) of $14.3$ K, sheet resistance of 12.4 $\Omega$/$\square$, and kinetic inductance of 1.2 pH/$\square$.

We vary the capacitor area by a factor of 44 on each chip with 20 resonators.
To keep the resonant frequency within the range of our setup, we reduce the length of the inductor for the larger capacitors, as shown in Fig.~\ref{fig:PPC_KIDS} and Fig.~\ref{fig:PPC_area}.
We vary the PPC thickness by fabricating the same design with three different dielectric thicknesses  of \textit{a}-$\mathrm{SiC}$:$\mathrm{H}$: 100, 200, and 400 nm.
The designed resonance frequencies on the 200-nm wafer lie between 4.0 and 8.45 GHz and will shift down or up for the 100- and 400-nm wafers, respectively.
For the 400-nm design, four resonators shifted beyond 9 GHz and could not be measured with our current setup.

We fabricate the resonators on a 350-$\mu$m \textit{c}-plane sapphire substrate using optical contact lithography and reactive-ion etching to pattern the structures.
The bottom PPC plate, inductor, readout lines and the bottom line of the coupling structure are fabricated from a 220-nm-thick reactive-magnetron sputtered layer of $\mathrm{NbTiN}$ \cite{Thoen2017,Bos2017}.
The $\mathrm{NbTiN}$ layer is etched with $\rm{O}_2$-$\rm{SF}_6$ (25 and 13.5 sccm, respectively) at 5 mTorr and 50 W for 460 s, with an additional 45-s overetch on the substrate to ensure a "sharp" definition of the structures.
We use an $\rm{O}_2$ (100 sccm, 100 mTorr, 50 W) etch of 90 s to remove any resist remnants.
The next step is a layer of \textit{a}-$\mathrm{SiC}$:$\mathrm{H}$ deposited by PECVD using a Novellus Concept One. 
Details regarding the \textit{a}-$\mathrm{SiC}$:$\mathrm{H}$ deposition are discussed in Ref.~ \cite{Buijtendorp2022}.
The top PPC plate, readout line bridges, and the top line of the coupling structure are fabricated from a second layer of 15-nm $\mathrm{NbTiN}$.
To create good galvanic contact between the two $\mathrm{NbTiN}$ layers, the wafer is cleaned with $\mathrm{HF}$ before the second $\mathrm{NbTiN}$ layer is deposited.
The use of a sapphire substrate prevents surface erosion due to the 45-s overetch needed for patterning the $\mathrm{NbTiN}$ and \textit{a}-$\mathrm{SiC}$:$\mathrm{H}$ layers.

In addition to the PPC resonators, we fabricate a set of planar reference devices in the bottom $\mathrm{NbTiN}$ layer.
These reference devices are used to separate the noise contributions from the capacitor and inductor \cite{McRae2020} which we come back to in the Sec.~\ref{sec:discussion}.

The samples are cooled in a pulse-tube precooled dilution refrigerator.
We use a box-in-box sample-stage design \cite{deVisser2014,Baselmans2012} to shield the sample from stray light coming from the 3-K stage of the cooler.
The sample stage is surrounded by two magnetic shields, a superconducting niobium shield, and a Cryophy shield.
A graphical representation of the setup is presented in Ref.~\cite{Kouwenhoven2023}, but the sample holder is closed with a lid.
Microwave measurements are performed with the standard homodyne detection scheme discussed in Appendix~\ref{sec:readout}.
Unless otherwise specified, all measurements are taken at a base temperature of 100 mK.


\section{Measurements}
\subsection{$S_{21}$ measurements}
\label{sec:S21}

\begin{figure*}[t!]
  \includegraphics{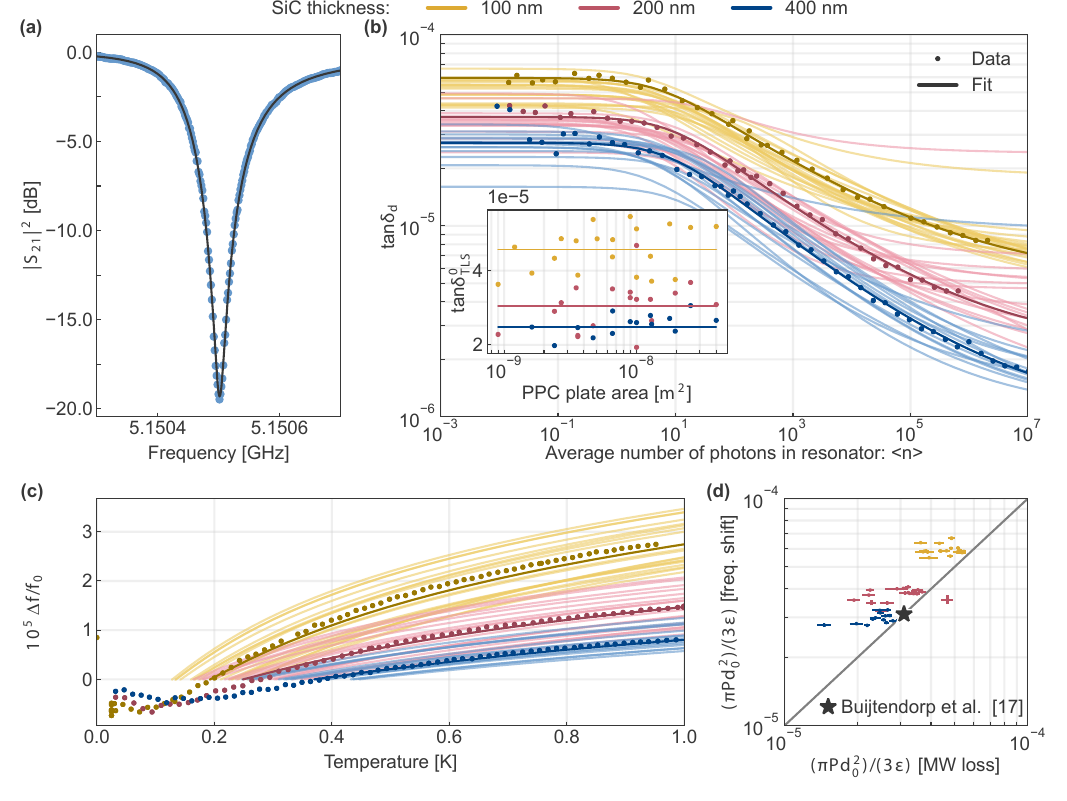}
  \caption{\textbf{(a)} Example resonance $|S_{21}|^2$ dip fitted in magnitude space with equations from Refs.~\cite{Khalil2012} and \cite{Swenson2013} for asymmetric nonlinear dips. \textbf{(b)} Power-dependent microwave loss $\tan\delta_d$, plotted against the average number of photons in the resonator ($\left<n\right>\propto{|\vec{E}|^2}$), with $\vec{E}$ as the electric field in the dielectric. Measured data points and the fit to Eq.~\ref{eq:TLS_loss} are highlighted for one KID of each thickness. For all other KIDs, only the fitted lines are presented. Error bars on each data point are small compared to the size of the data points. \textbf{Inset} shows that the TLS contribution to the microwave loss, $\tan\delta^0_{\rm{TLS}}=(\pi Pd_0^2)/(3\epsilon)$, is independent of the parallel-plate area. Solid lines are the average $\tan\delta^0_{\rm{TLS}}$ of each thickness, presented as a guide to the eye (not a fit). \textbf{(c)} Temperature-induced relative frequency shift. Measured data points for three KIDs, together with the fit to Eq.~\ref{eq:TLS_freq}, are highlighted for each dielectric thickness. For all other KIDs only the fitted lines are presented. Fitted range is $\hbar\omega/k_{B}$ - $1 \rm{K}$ which, for a KID resonating at 4 GHz starts at about 200 mK. At $\hbar\omega=k_{B}T$, $\delta f/f_0$ crosses zero. Error bars on each data point are small compared to the size of the data points. \textbf{d)} Fitted TLS contribution $(\pi Pd_0^2)/(3\epsilon)$. \ding{72} aata point from a 295 nm \textit{a}-$\mathrm{SiC}$:$\mathrm{H}$ microstrip resonator with identical \textit{a}-$\mathrm{SiC}$:$\mathrm{H}$ deposition conditions \cite{Buijtendorp2022}; $(\pi Pd_0^2)/(3\epsilon)$ obtained from the loss tangent fit. TLS contribution is thickness dependent, which cannot be explained by a uniform dielectric that assumes a constant TLS density ($P$) and dipole moment ($d_0$).}
  \label{fig:S21_results}
\end{figure*}

We measure the complex transmission parameter, $S_{21}$, with a vector network analyzer (VNA) to characterize the resonators. 
We perform a frequency sweep to obtain each resonator's resonance circle ($S_{21}$) at different readout powers and sample-stage temperatures.

Fitting the resonance dip in magnitude space $(|S_{21}|)$ to an analytical model \cite{Khalil2012} provides the resonator's resonance frequency, $f_r$, and the internal ($Q_i$) and coupling ($Q_c$) quality factors as a function of readout power and temperature.
At the highest readout powers the resonance dip ($|S_{21}|$) becomes nonlinear \cite{Swenson2013}.
To estimate the $Q$ factors at these powers we include the nonlinear frequency response \cite{Swenson2013} in the analytical model of Ref.~\cite{Khalil2012}.
The maximum readout power we use is set by the point at which the analytical model that combines the resonance dip asymmetry \cite{Khalil2012} and  nonlinearity \cite{Swenson2013} no longer represents the observed resonance dip ($|S_{21}|$).

The resonator's intrinsic loss is given by $Q_i$, which is related to the loss tangent of the dielectric,

\begin{equation}
	\tan\delta_d = \frac{1}{pQ_i},
	\label{eq:part_ratio}
\end{equation}

if the dielectric material dominates the resonator's microwave loss.
Here, $p$ is the participation ratio of the dielectric, given by $p=w^e_d/w^e$, where $w_e$ is the total electric energy stored in the resonator and $w^e_d$ is the energy stored in the dielectric.
For a PPC where the dielectric thickness is small compared to the dimensions of the plates, the fringing fields are negligible, so we can assume that the entire electric field is in the dielectric and $p=1$.

We measure $Q_i$ over a wide power range, from the bifurcation point of the resonator down to the single-photon level.
The measured data points of three resonators, one for each PPC thickness, are presented in Fig.~\ref{fig:S21_results}(a).
The STM predicts a power-dependent microwave loss:

\begin{equation}
	{\tan\delta_d} =
	\frac{\pi Pd_0^2}{3\epsilon}\tanh{\frac{\hbar\omega_r}{2k_{B}T}}
	\left(1 + \frac{{|\vec{E}|}}{{E_c}} \right)^{-\beta} + 
	{\tan\delta_{\rm{HP}}},
	\label{eq:TLS_loss}
\end{equation}

which should be geometry independent.
Here, $P$ is the TLS density, $d_0$ is the TLS dipole moment, $\epsilon$ is the dielectric constant of the TLS hosting material, ${|\vec{E}|}$ is the electric field in the dielectric, $\omega_r$ is the resonance frequency, $T$ is the temperature, and $E_{\rm{c}}$ is the critical field above which the TLSs start to saturate with $\beta = 0.5$ in the STM.
The TLS contribution, $(\pi Pd_0^2)/(3\epsilon)$, is often presented as the TLS contribution to the loss tangent at zero temperature: $\tan\delta^0_{\rm{TLS}}$.
The loss no longer follows the STM at high powers and saturates to $\tan\delta_{\rm{HP}}$.

Figure \ref{fig:3_loss}(b) shows the measured loss tangent ($1/Q_i$) versus the average number of microwave photons in the resonator,

\begin{equation}
	\left< n\right> = \frac{2Q^2}{Q_c}\frac{P_{\rm{read}}}{\hbar\omega_r^2},
\end{equation}

where $\langle n \rangle \propto|\vec{E}|^2$, $\omega_r$ is the resonance frequency, $Q$ is the loaded quality factor, and $Q_c$ is the coupling quality factor.
The readout power at the sample, $P_{\rm{read}}$, is calibrated by measuring the transmission through two identical input lines, including attenuation, with a short cable instead of the sample, see Appendix~\ref{sec:readout}.
To extract the TLS contribution to the microwave loss we, fit Eq.~\ref{eq:TLS_loss} to the measured internal quality factor, $Q_i$.
Fig.~\ref{fig:S21_results}(b) gives the fits for all resonators.
The fitted TLS contribution, $\tan\delta^0_{\rm{TLS}}$ ,of each resonator is plotted against the PPC area in the inset of Fig.~\ref{fig:S21_results}(b), which shows that $\tan\delta^0_{\rm{TLS}}$ is independent of the PPC area.

For the wafer with 200 nm of \textit{a}-$\mathrm{SiC}$:$\mathrm{H}$, we observe that some resonators become nonlinear at lower readout powers ($< -20$dB) compared to the other devices on the same chip \cite{Swenson2013}.
For these devices, there are not enough data points to accurately fit Eq.~\ref{eq:TLS_loss}, and they are therefore omitted from the analysis.
We measured a second chip of the same design and fabrication run to extend the dataset.
The full dataset, with all measured devices, is available in the reproduction package \cite{repr_zenodo}.

Next, we analyze the KID resonance frequency as a function of temperature by varying the temperature of the sample stage with a PID-controlled heater, from $25.0$ mK to $1.0$ K.
The STM predicts a temperature-dependent shift of the KID resonance frequency ($\Delta f_0 / f_0 = -\frac{1}{2}\Delta \epsilon'/\epsilon$):

\begin{equation}
	\frac{\Delta f_0}{f_0} =
	p\frac{Pd_0^2}{3\epsilon}
	\left[
	\mathrm{Re}\left\{ \Psi\left(\frac{1}{2} + \frac{\hbar\omega}{2\pi jk_{B}T}\right)\right\} -
	\log{\frac{\hbar\omega}{2\pi k_{B}T}}
	\right],
	\label{eq:TLS_freq}
\end{equation}

which should be geometry independent \cite{Gao2008}.
Here, $\Delta f_0 = f_r(T)-f_0$, where $f_0$ is the resonance frequency at $T=0$; $\Psi$ is the complex digamma function; and $p$ is the participation ratio, as in Eq.~\ref{eq:part_ratio}.
We fit Eq.~\ref{eq:TLS_freq} to the measured resonance frequencies, highlighted for one resonator of each dielectric thickness in Fig.~\ref{fig:S21_results}(c).
For all other resonators, only the fitted line is presented.

Equation~\ref{eq:TLS_freq} contains two regimes, for $k_{B}T<\hbar\omega$ the digamma term dominates, while, for $k_{B}T>\hbar\omega$ the logarithmic term dominates.
In the first regime ($k_{B}T<\hbar\omega$) we have a low signal-to-noise ratio and expect TLS saturation effects \cite{Gao2008}, so we limit the fit to the range $k_{B}T>\hbar\omega$.
Since the measured temperature range is far below the $T_{\rm{c}}$ of $\mathrm{NbTiN}$ (14.3 K), there is no quasiparticle contribution to the measured frequency shift \cite{Mattis1958,Barends2008}.

The only TLS parameters that impact the TLS-induced microwave loss and resonance frequency shift are the TLS density ($P$) and the TLS dipole moment ($d_0$).
The fitted TLS contributions, $(\pi Pd_0^2)/(3\epsilon)$, from both Eq.~\ref{eq:TLS_loss} and Eq.~\ref{eq:TLS_freq}, for all devices, are plotted in Fig.~\ref{fig:S21_results}(d).
There is a clear close to 1:1 agreement between the obtained TLS contribution, $(\pi Pd_0^2)/(3\epsilon)$ ,from both measurements.

The TLS data in Fig.~\ref{fig:S21_results} show a clear thickness dependence, where a thicker dielectric has both a lower TLS-induced microwave loss [Fig.~\ref{fig:S21_results}(b)] and a smaller TLS-induced resonance frequency shift [Fig.~\ref{fig:S21_results}(c], as summarized by Fig.~\ref{fig:S21_results}(d).
The inset of Fig.~\ref{fig:S21_results}(b) highlights that the TLS-induced microwave loss does not depend on the area of the PPC and is thus only a function of the thickness of the dielectric.
The thickness dependence cannot be explained by a uniform dielectric, which assumes a constant TLS density ($P$) and dipole moment ($d_0$) in Eq.~\ref{eq:TLS_loss} and Eq.~\ref{eq:TLS_freq}.
Instead, the medium between the PPC plates likely contains one or more surface layers with a higher TLS density or a stronger TLS dipole moment (or both) than the bulk dielectric.
The TLS contribution we obtain in Fig.~\ref{fig:S21_results} is then the thickness-weighted average of the bulk dielectric material and surface layers.
If the thickness of the dielectric changes, the bulk-to-surface-layer ratio changes, which yields a different effective TLS contribution, $(\pi Pd_0^2)/(3\epsilon)$.

\subsection{Noise measurements}
\label{sec:Noise}

\begin{figure*}[t!]
  \includegraphics{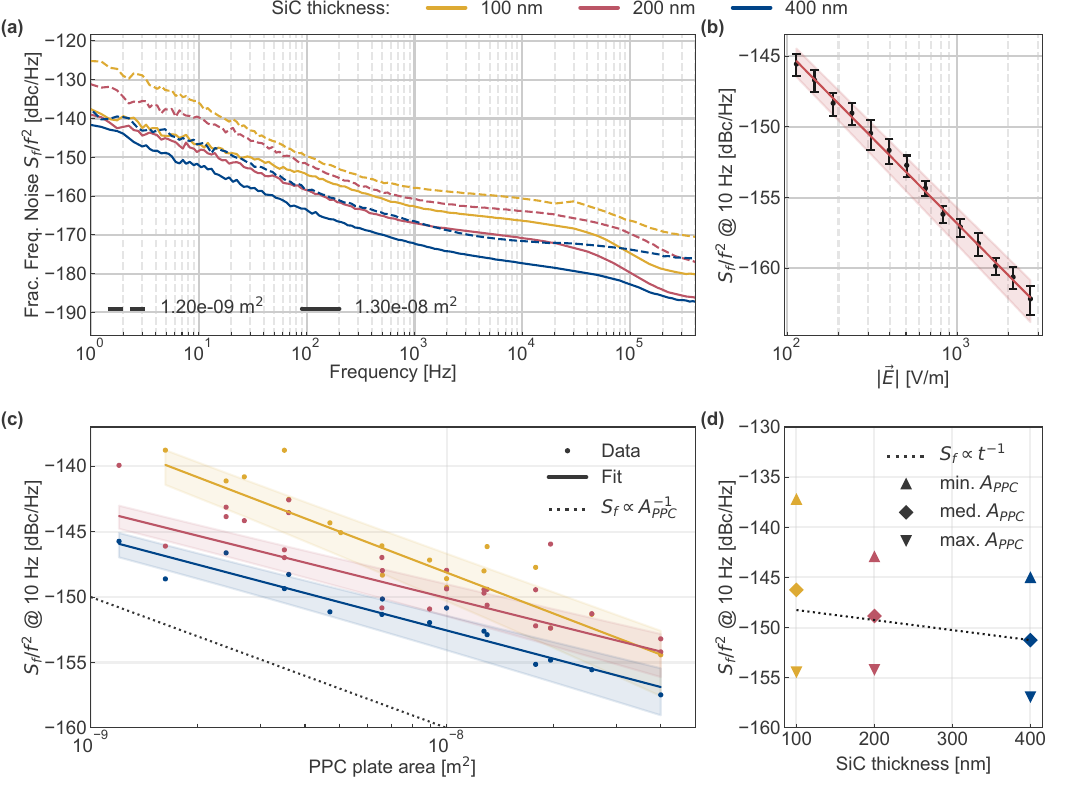}
  \caption{\textbf{a)} Frequency noise spectra for resonators with different plate sizes and dielectric thickness, at the same internal electric field of $|\vec{E}|=1.13\times10^3$ V/m ($\epsilon_r=10$). TLS noise level shows a clear area and thickness scaling. Spectra typically show a 1/$f$ spectrum up to 0.1-1 kHz. Roll-off in the noise level at the resonator-ring time is visible at higher frequencies ($>10^4$ Hz). \textbf{b)} Frequency noise level of one PPC resonator ($A = 200\times200$ \SI{}{\micro\meter\squared}, $t = 400$ nm, $f_r = 5.15$ GHz) as a function of $|\vec{E}|$, which follows the expected $|\vec{E}|^{-1}$ scaling from the tunneling models \cite{Gao2008thesis, Faoro2015}. \textbf{c)} TLS noise level at 10 Hz versus parallel-plate area ($A_{\rm{PPC}}$) for all three \textit{a}-$\mathrm{SiC}$:$\mathrm{H}$ thicknesses at the same internal electric field. Solid lines are fits $cA_{\rm{PPC}}^{-\alpha}$. From thinnest to thickest dielectric layer, the slopes ($\alpha$) are $1.04 \pm 0.12$, $0.73 \pm 0.10$, and $0.72 \pm 0.09$. Standard-deviation error on the fit parameters $\alpha$ and $c$ gives the uncertainty range. Dotted line indicates the slope of the expected $S_f\propto A_{\rm{PPC}}^{-1}$ scaling. \textbf{d)} Observed thickness scaling of the TLS noise level. Points are then obtained from the fitted $S_f(A_{\rm{PPC}})/f^2$ relations in (c), at the smallest PPC area (\ding{115}), the median area of $6.6\times10^{-9}$ $\rm{m}^2$ (\ding{117}), and the largest PPC area (\ding{116}). Dashed line indicates a $t^{-1}$ scaling with respect to the 400-nm median-area point.}
  \label{fig:TLS_noise}
\end{figure*}

We perform noise measurements at the resonator's resonance frequency with the homodyne detection scheme discussed in Appendix~\ref{sec:readout}.
The noise measurement consists of two timestreams: one of 40 s sampled at 50 ksample/s and one of 0.5 s sampled at 1 Msample/s.
The fluctuations in the IQ coordinates are translated to phase ($\theta$) and amplitude ($A$) with respect to the resonance circle's center \cite{deVisserthesis}.
We estimate the PSD from both timestreams and stitch them at 20 kHz.
The resonance circle relates the measured phase noise to frequency noise as 

\begin{equation}
	S_f/f_r^2 = \frac{S_{\theta}}{(4Q_{l})^2}
	\label{eq:freq_noise}
\end{equation}

where $S_{\theta}$ is the PSD in the phase coordinate.
The resonance frequency, $f_r$, and loaded quality factor, $Q_{l}=(1/Q_{c} + 1/Q_i)^{-1}$, are obtained from the resonance circle, see sec.~\ref{sec:S21}.
We focus on the frequency noise from the phase coordinate, which has a higher TLS noise level compared to the amplitude coordinate at powers where $|\vec{E}|>E_{\rm{c}}$ \cite{Neill2013}.

The frequency noise spectrum for a microwave resonator, for which the total electric field volume, $V$, contains a TLS host volume, $V_h$, is given by \cite{Gao2008_semiemperical, Gao2008thesis}

\begin{equation}
	\frac{S_{f}(f)}{f^2} \sim 
	\frac{\int_{V_h}S_\epsilon(f,\vec{E},T)|\vec{E}|^{4}dV}
	{4\left(\int_{V}\epsilon|\vec{E}|^{2}dV\right)^{2}}
	\label{eq:noise_scaling}
\end{equation}

where $S_\epsilon$ describes the fluctuations in the real part of the dielectric constant ($\epsilon_{\rm{TLS}}$) due to random switching of TLSs.
If we assume a uniform distribution of TLS in the dielectric volume of the PPC $V_{\rm{PPC}}$, then $V_h = V = V_{\rm{PPC}}$  and the frequency noise level should scale as $V_{\rm{PPC}}^{-1}$.
The derivation for the interacting tunneling model \cite{Faoro2015} shows that $S_\epsilon$ has a $1/f$ spectrum and in strong electric fields ($|\vec{E}|\gg E_{\rm{c}}$) scales with $|\vec{E}|^{-1}$, matching the empirical relations of the STM \cite{Gao2008_semiemperical, Gao2008thesis}.
The interacting TLSs do, however, result in a temperature-dependent noise level that is not explained by the STM: at low temperatures ($k_{B} T \ll \hbar\omega$) and strong field ($|\vec{E}|\gg E_{\rm{c}}$) $S_f \propto T^{(1-\mu)/2}$, while, at high temperatures ($k_{B} T \gg \hbar\omega$) $S_f \propto T^{\mu-1}$, where $\mu\approx0.3$.

The measured frequency spectra ($S_{f}(f)/f^2$) of six KIDs, with different PPC areas and dielectric thicknesses, are presented in Fig.~\ref{fig:TLS_noise}(a) at a temperature of 100 mK.

The frequency noise spectra show apparent $1/f$ behavior at lower frequencies, transitioning to something more akin to $f^{-1/2}$, typically between 0.1 and1 kHz.
Both behaviors have been observed before in different measurements and device architectures \cite{Gao2007, Barends2008, Gao2008_semiemperical, Burnett2014}, but typically not simultaneously.
The transition frequency varies between resonators and geometries, see Appendix.~\ref{sec:ref_devices}.
To compare the noise levels of different resonators, we fit the $1/f$ spectrum between 5 and 50 Hz, where the spectrum is always fully $1/f$, and use the fitted 10-Hz point as the TLS noise-level reference.

We measure the noise spectra over a range of readout powers with 2-dB steps.
In general, the electric field in the TLS medium is proportional to the readout power on the through line as $|\vec{E}| \propto P_{\rm{read}}^{-1/2}$. 
For a PPC, the internal electric field is directly calculated from the readout power and geometry as

\begin{equation}
\left|\vec{E}\right|=2\sqrt{\frac{\pi P_{\rm{int}}}{\omega_r\epsilon}\frac{1}{V_{\rm{PPC}}}},
\label{eq:E-field}
\end{equation}

where $\omega_r$ is the resonance frequency, $\epsilon$ is the dielectric constant, and $V_{\rm{PPC}}$ is the volume of the dielectric in the PPC.
The internal power, $P_{\rm{int}}$, is given by $P_{\rm{int}}=Q^2 / \left(\pi Q_c \right)P_{\rm{read}} = \langle n \rangle\hbar\omega^2 / \left(2\pi\right)$.
The frequency noise level at 10 Hz for one PPC resonator ($A = 200\times200$ \SI{}{\micro\meter\squared}, $t = 400$ nm, $f_r = 5.15$ GHz) for different electric fields is given in Fig.~\ref{fig:TLS_noise}(b) to illustrate that we indeed observe the expected $\propto|\vec{E}|^{-1}$ relationship.

In addition, we have measured the temperature dependence of the frequency noise.
The frequency noise spectra and noise levels at 10 Hz for temperatures between $25 and 800$ mK, and at different internal powers, are discussed in Appendix~\ref{sec:Temp_dep} and plotted in Fig.~\ref{fig:Noise_T}.
At all measured powers, where $|\vec{E}| \gg E_c$, the frequency noise level follows the temperature dependence predicted by the interacting tunneling model.

Neither the STM nor the interacting tunneling model \cite{Faoro2015} predict the $f^{-1/2}$ region of the noise spectra we typically observe for superconducting resonators \cite{Gao2007, Barends2008, Gao2008_semiemperical}.
The $f^{-1/2}$ region roughly follows the expected $|\vec{E}|$ and temperature scalings of the $1/f$ region \cite{Faoro2015}, which suggests that the $f^{-1/2}$ region has a similar field and temperature dependence to the TLSs and is therefore interesting for follow up.


For a uniform dielectric where the TLSs are uniformly distributed over the dielectric volume, we expect a frequency noise level that scales as $S_f \propto V_{\rm{PPC}}^{-1}$.
Separated in the two geometrical parameters of the PPC, PPC area and thickness, we expect that the frequency noise level scales the same with area, $S_f \propto A_{PPC}^{-1}$, and with thickness, $S_f\propto t^{-1}$.

To see the effect of the PPC dimensions on the TLS noise level, we need to compare the noise levels at equal electric fields in the resonator.
For each resonator, we translate the measured read power points to the electric field in the dielectric (Eq.~\ref{eq:E-field}).
We interpolate between these measured electric field points, using the known $S_f \propto |\vec{E}|^{-1}$ relationship, to find the noise level at a desired electric field, see Fig.~\ref{fig:TLS_noise}(b).
Details of the interpolation are discussed in Appendix \ref{sec:freq_noise_a}.
The TLS noise level, at 10 Hz, versus the PPC plate area at $|\vec{E}|=1.13\times10^3$ V/m, assuming $\epsilon_r=10$ for \textit{a}-$\mathrm{SiC}$:$\mathrm{H}$, is given in Fig.~\ref{fig:TLS_noise}(c).
 
As expected, the frequency noise level in Fig.~\ref{fig:TLS_noise}(c) scales as a power law with the parallel-plate area.
 We fit the relation $S_f=\beta A_{\rm{PPC}}^{-\alpha}$, where $A_{\rm{PPC}}$ is the PPC area, and find that $\alpha$ is $1.04\pm0.12$ for 100 nm, $0.73\pm0.10$ for 200 nm, and $0.72\pm0.09$ for 400 nm.
 The STM predicts $S_f \propto A_{\rm{PPC}}^{-1}$, as in Eq.~\ref{eq:noise_scaling}.
The obtained area scalings qualitatively agree with the predicted scaling for a uniform dielectric.
However, we expect all three dielectric thicknesses to have the same frequency noise scaling with area.
Instead, we observe that the frequency noise scaling is stronger with area for the 100-nm thick film than for the 200- and 400-nm films.


 The resonance frequency scaling, $S_f \propto f_0^\mu$, from the interacting tunneling model \cite{Faoro2015}, with $\mu\approx 0.3$, would only introduce a scatter on the data points in Fig.~\ref{fig:TLS_noise}(c), since the resonance frequency is, by design, not correlated with the PPC area.
 
In Fig.~\ref{fig:TLS_noise}(d), we compare the fitted noise level for the different dielectric thicknesses.
Since the slopes in Fig.~\ref{fig:TLS_noise}(c) differ, we compare the noise level at different PPC surface areas: the smallest area (\ding{115}), the median area (\ding{117}), and the largest area (\ding{116}).
If we compare data from the 200- and 400-nm films, which show a similar area scaling ($\sim A_{\rm{PPC}}^{-0.7}$), we see that the frequency noise level scales roughly with the expected $t_{\rm{PPC}}^{-1}$ scaling for a uniform dielectric.

In Sec.~\ref{sec:S21}, we concluded that the dielectric contained surface layers based on the thickness-dependent microwave loss and resonance frequency shift.
If these surface layers have a similar effect on the frequency noise, we expect a frequency noise scaling with thickness stronger than the scaling for a uniform dielectric ($S_f\propto t^{-1}$).
The data in Fig.~\ref{fig:TLS_noise}d do not show a thickness dependence that is unambiguously stronger than $S_f\propto t^{-1}$, which implies that the surface layers affect the noise differently than the microwave loss and frequency shift discussed in Sec.~\ref{sec:S21}.
The origin of this difference is unknown.
Barends et al. \cite{Barends2008} made a similar observation: the resonance frequency shift and frequency noise respond differently to thicker and thicker dielectric ($\rm{SiO_x}$) layers deposited on top of a CPW resonator.

\section{Discussion and Conclusions}
\label{sec:discussion}

\begin{figure}[hbt]
  \includegraphics{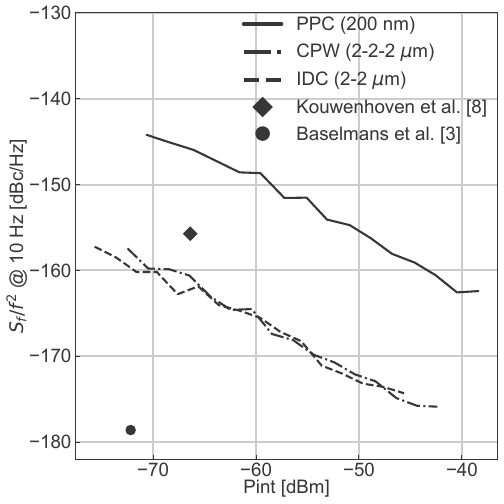}
  \caption{TLS noise level at 10 Hz versus internal power for different resonator types. PPC, IDC, and CPW resonators are fabricated on the same chip, where the PPC and IDC share the same inductor design. PPC has a plate area of 6000 \SI{}{\micro\meter\squared} (60$\times$100 \SI{}{\micro\meter\squared}). There is a strong increase in noise level between the planar devices deposited on the crystalline substrate (CPW, IDC) and the devices based on the \textit{a}-$\mathrm{SiC}$:$\mathrm{H}$ parallel-plate capacitor, but no difference between the planar devices themselves. Markers are the frequency noise of an optical-to-near-IR energy-resolving KID (pixel pitch 150 $\mu$m) \cite{Kouwenhoven2023} and an ultra-sensitive far-infrared KID (pixel pitch $\sim$1.5 mm) \cite{Baselmans2022}, both operated at their highest internal power.} 
  \label{fig:device_noise}
\end{figure}

In addition to the PPC designs, each wafer contains single-layer reference devices: CPW and IDC resonators.
The measured microwave loss and frequency noise for these devices are presented in Appendix.~\ref{sec:ref_devices}.
The frequency noise against internal power is plotted in Fig.~\ref{fig:device_noise}.
Calculating the electric field inside the TLS medium for a CPW or IDC is nontrivial due to the degeneracy in $p\tan{\delta_d}$ (Eq.~\ref{eq:part_ratio}).
This means we cannot quantitatively compare the TLS properties of the substrate to \textit{a}-$\mathrm{SiC}$:$\mathrm{H}$, but we can show that the PPC resonator is dominated by a different TLS contribution than the CPW and IDC resonators.

The inductor's loss and frequency noise contribution can be analyzed by comparing the IDC and CPW resonators, which have an equal line and gap width, and thus, the same field distribution in the substrate TLS layer \cite{McRae2020}.
The difference between the CPW and IDC resonators is the inductor in the IDC resonator.
Figure~\ref{fig:device_noise} shows that the CPW and IDC have similar frequency noise levels, which means the contribution of the inductor is negligible.
The IDC and PPC resonators have the same inductor but a different lumped-element capacitor.
Compared to the IDC designs, the PPC resonators show a 15--20-dB-higher TLS noise level in the $1/f$ region of the spectrum.
This means the noise of the PPC resonators is dominated by the \textit{a}-$\mathrm{SiC}$:$\mathrm{H}$ capacitor.

In the measured microwave loss and resonance frequencies in Fig.~\ref{fig:S21_results}, we observe a thickness dependency that can be explained by a volume between the PPC plates that is not one uniform dielectric bulk material but contains one or more surface layers.
Since these surface layers have a different $\tan\delta_{\rm{TLS}}$ than the bulk, they could affect the frequency noise of the PPC as well, where the noise level scales more strongly with the thickness of the dielectric than with the PPC area.
The measured frequency noise in Fig.~\ref{fig:TLS_noise} does not clearly indicate that we see this effect, see Sec.~\ref{sec:Noise}.
This could mean that the surface layers affect the noise differently than the microwave loss and frequency shift.

The location or nature of the surface layer(s) is unknown.
They can be surface oxides on the metal capacitor plates or growth effects of \textit{a}-$\mathrm{SiC}$:$\mathrm{H}$.
To find $\tan\delta_{\rm{TLS}}$ of the bulk dielectric, we would need to extend the dielectric thickness range far enough to reach two extremes: thin enough to be dominated by the surface layers and thick enough to be limited by the properties of the bulk dielectric.
A way to mitigate these surface layers might be found by exploring various cleaning steps on the bottom $\mathrm{NbTiN}$ plate of the capacitor.


One of the applications of superconducting resonators is the KID.
Since the noise level scales with $P_{\rm{int}}$, KIDs are always operated at their maximum internal power, $P_{\rm{int}}^{\rm{max}}$, at the edge of bifurcation \cite{deVisser2010,Swenson2013}.
The bifurcation point, and thus, $P_{\rm{int,max}}$, typically depends on the critical current in the narrow inductor lines and does not depend on the capacitor geometry.
For a set $P_{\rm{int,max}}$, the internal electric field inside the capacitor will depend on the capacitor geometry, see Eq.~\ref{eq:E-field}.
This gives us two PPC dimension scalings, one for the internal electric field and one for the dielectric volume ($V_{\rm{PPC}}$) in the resonator, as discussed before:

\begin{equation}
	S_f \propto
	\frac{1}{\vec{|E|}}
	\frac{1}{V_{\rm{PPC}}},
	\label{eq:S_T}
\end{equation}

where $|\vec{E}|\propto 1/\sqrt{\omega_r V_{\rm{PPC}}}$.
The two dielectric volumes factors, $V_{\rm{PPC}}$, partly cancel and a 

\begin{equation}
	S_f \propto \sqrt{\omega_r/V_{\rm{PPC}}}
	\label{eq:practical_scaling}
\end{equation}

scaling remains for a dielectric with uniform TLS density.
However, if we look at the measured data in Fig.~\ref{fig:TLS_noise}, we see that, for the 200- and 400-nm-thick films, we have a geometric scaling

\begin{equation}
	S_f \propto A_{\rm{PPC}}^{-\alpha} t^{-\beta} < V_{\rm{PPC}}^{-1},
\end{equation}

since $\alpha \approx 0.72 - 0.73$ (see Fig.~\ref{fig:TLS_noise}(c)) and $\beta \approx 1$ (see Fig.~\ref{fig:TLS_noise}(d)).
The resulting geometric volume scaling, which is weaker than $V_{\rm{PPC}}^{-1}$, reduces the combined geometry and electric field scaling in Eq.~\ref{eq:practical_scaling}.

The practical applicability of PPC resonators for KIDs depends on how their noise level compares to the standard planar structures.
Figure~\ref{fig:device_noise} shows the measured $S_f/f^2$ of a 200-nm PPC KID and of the planar reference designs fabricated on the same chip.
In addition, Fig.~\ref{fig:device_noise} contains data points from two different MKID designs operated at their maximal internal power.
The first is a compact lumped-element KID (LEKID) design for an optical-to-near-IR energy-resolving pixel based on an IDC \cite{Kouwenhoven2023}.
The second point is for ultra-sensitive antenna-coupled terahertz KIDs based on planar structures \cite{Baselmans2022}.

%

Compared to the IDC of the LEKIDs in Ref. \cite{Kouwenhoven2023}, which have 2-$\mu$m fingers and gaps, the 200-nm PPC of Fig.~\ref{fig:device_noise} with 60$\times$100-$\mu$m sides has a capacitance that is 42 times higher.
This means we can replace the IDC of Ref. \cite{Kouwenhoven2023} with a 12$\times$12 \SI{}{\micro\meter\squared} PPC to get the same KID resonance frequencies.
Assuming the noise scaling from Eq.~\ref{eq:practical_scaling}, this would add roughly 8 dB of frequency noise to the 200-nm PPC line in Fig.~\ref{fig:device_noise}.
Considering the inductor's dimensions, we can use the \textit{a}-$\mathrm{SiC}$:$\mathrm{H}$ PPC to reduce the area required for a KID by a factor of 10 at the cost of roughly 20 dB of extra frequency noise.

The second application is in ultra-sensitive KIDs for the far-infrared (FIR), where one of the limiting factors is the frequency noise at low frequencies ($0.1 - 10$ Hz) \cite{Baselmans2022}.
The FIR KIDs \cite{Baselmans2022} have a TLS noise level far below both the PPC and the reference resonators, see Fig.~\ref{fig:device_noise}.
The noise level of these KIDs is so low because they are based on wide planar structures (central line of 40 $\mu$m with 8-$\mu$m gaps) that leverage the $W_{tot}^{-1.6}$ scaling of the frequency noise \cite{Gao2008_semiemperical, Hahnle2021}.
Compared to the expected $V_{\rm{PPC}}^{-1/2}$ scaling for a PPC resonator with a uniform dielectric, the CPW has a much stronger geometrical scaling.
This means that PPC-based resonators are at a disadvantage for FIR KID arrays, which are not limited by the pixel pitch ($\sim$ 1.5 mm).
A promising route would be to investigate the dielectric material's properties and eliminate any possible TLS surface layers to reduce the frequency noise level of a PPC-based resonator.

\section*{Data availability}
The full dataset of all measured resonators, and the reproduction package that generates the figures are available on Zenodo: \url{https://doi.org/10.5281/zenodo.10159731}.
The colorblind and grayscale safe color scheme is from Paul Tol's Notes: \url{https://personal.sron.nl/~pault/}

\section*{Acknowledgments}
We acknowledge Nick de Keijzer and Robert Huiting for their work on the 100 mK sample stage.
This work is financially supported by the Netherlands Organisation for Scientific Research NWO (Projectruimte 680-91-127)

B. T. Buijtendorp is supported by the European Union (ERC Consolidator Grant No. 101043486 TIFUUN). Views and opinions expressed are however those of the authors only and do not necessarily reflect those of the European Union or the European Research Council Executive Agency. Neither the European Union nor the granting authority can be held responsible for them.

\appendix{}

\section{Area variation}
\label{sec:Area_var}

\begin{figure}[hbt]
	\includegraphics{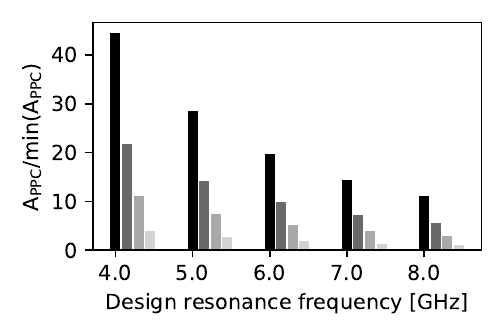}
  	\caption{The area of each of the 20 PPC resonators with respect to the smallest PPC plotted against their designed resonance frequency for the 200 nm \textit{a}-$\mathrm{SiC}$:$\mathrm{H}$ wafer. Bars of the same color correspond to resonators with identical inductor lengths; black corresponds to the shortest inductors. There is no direct correlation between PPC area and resonance frequency.}
  	\label{fig:PPC_area}
\end{figure}

The area of each of the 20 PPC resonators with respect
to the smallest PPC is plotted against their designed
resonance frequency for the 200-nm \textit{a}-$\mathrm{SiC}$:$\mathrm{H}$ wafer in
Fig.~\ref{fig:PPC_area}.

\section{Measurement setup}
\label{sec:readout}

\begin{figure}[hbt]
	\includegraphics[width = \linewidth]{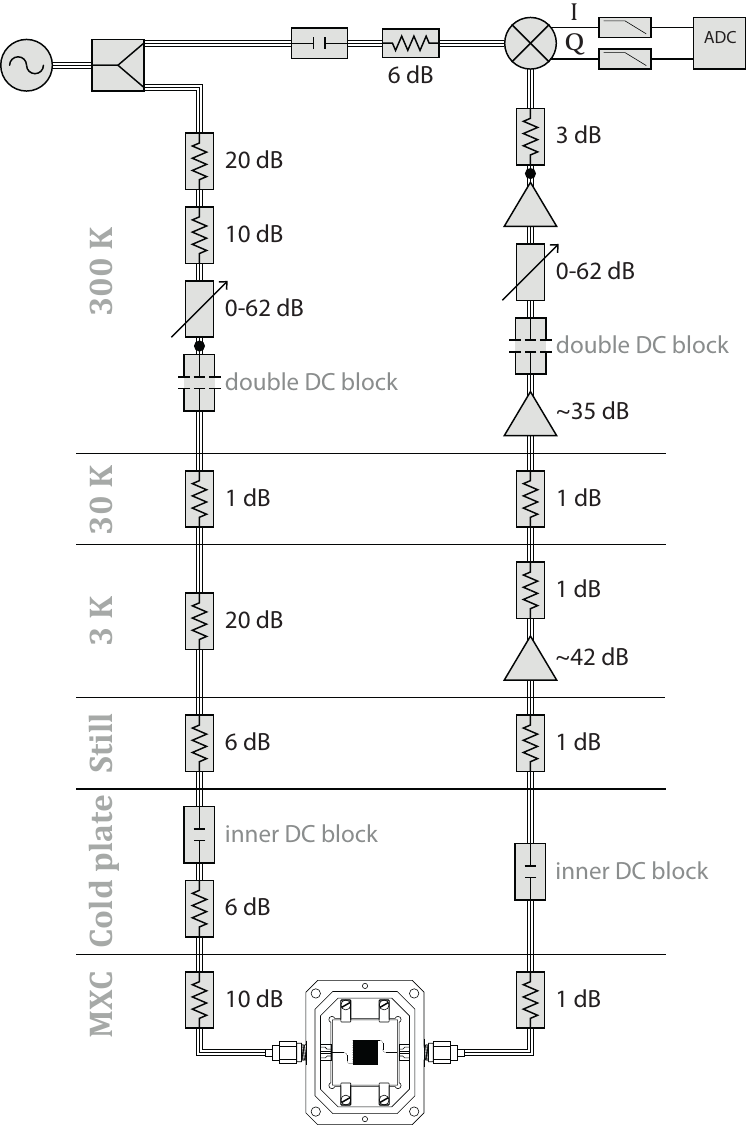}
  	\caption{Microwave components in the readout system. Room-temperature coaxial cables are $\varnothing$3.58-mm copper. At 30 K, the cables are $\varnothing$2.19-mm steel. At the input, the cables from the 20-dB attenuator up to the 1-dB attenuator at the mixing chamber (MXC)\textemdash or ``sample stage''\textemdash are $\varnothing$0.86-mm $\mathrm{CuNi}$. Between the two 1-dB attenuators, the cable is $\varnothing$2.30-mm $\mathrm{Al}$, and before the first amplifier $\varnothing$0.86-mm $\mathrm{NbTi}$. Power at the sample is set by the adjustable attenuators (Weinschel 8310). For measurements between 4 and 8 GHz, the amplifier at the 3-K stage is a Low Noise Factory LNF-LNC4\_8C and the warm amplifier at 300 K is a MITEQ LNA-30-04000800-07-10P. Alternatively, they can be replaced by a Low Noise Factory LNF-LNC2\_6A and a MITEQ LNA-30-02000600-09-10P for measurements between 2 and 6 GHz. For the $S_{21}$ measurements, the section above the black dots is replaced by a VNA (Keysight N5230A PNA-L).}
  	\label{fig:readout}
\end{figure}

For the noise measurements, we use the homodyne readout presented in Fig.~\ref{fig:readout}.
The signal from the generator (Agilent E8257D) is split in two. 
One part is sent directly to the IQ demodulator (MITEQ IRM0218LC1Q) as a reference signal, while the other part passes through the readout sample where it is influenced by the complex transmission, $S_{21}$, of the resonator.
The output of the IQ demodulator is digitized by an analog-to-digital converter (National Instruments PXI-5922).

Before reaching the sample, the input microwave signal is first attenuated at every temperature stage.
The power at the sample is set by the adjustable attenuators (Weinschel 8310).
After the sample, the signal is amplified with a set of low-noise amplifiers (LNAs), of which one is located at the 3-K stage.
The first amplifier (Low Noise Factory LNF-LNC4\_8C) is the dominant contribution to the noise figure of the system.
At 3 K, this amplifier has an equivalent noise temperature of 1.5 K, which, due to the two attenuators at 100 mK and 1, results in a 2.7-K equivalent system noise temperature \cite{barendsthesis, deVisserthesis}.

For the $S_{21}$ measurements the parts above the thick black dots in Fig.~\ref{fig:readout} are replaced by a VNA (Keysight N5230A PNA-L).
To reach the low powers required for the loss tangent measurements in Fig.~\ref{fig:3_loss}, the signal is attenuated before entering the cryostat and amplified after the second LNA by an extra 37 dB.

\section{Reference devices (CPW and IDC)}
\label{sec:ref_devices}

\begin{figure}[hbt]
  \includegraphics{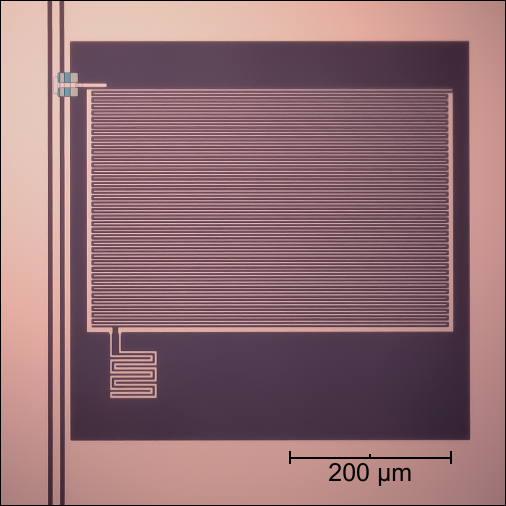}
  \caption{Microscope image of one of the IDC reference resonators. IDC has 2 $\mu$m defined fingers and gaps. Single-layer planar resonator fabricated in the bottom $\mathrm{NbTiN}$ layer which has a $T_c$ of 14.3 K.}
  \label{fig:IDC}
\end{figure}

In addition to the parallel-plate capacitor designs, each wafer contains reference designs based on the work in Ref. \cite{McRae2020}.
In addition to three PPC resonators, this chip contains three IDC resonators and three CPW resonators.
A microscope image from one of the IDC resonators is presented in Fig. \ref{fig:IDC}.
The IDC has fingers and gaps of 2 $\mu$m, and the CPW has a center line and gap width of 2 $\mu$m (2-2-2) with same electric field distribution.
Both resonators should then probe the TLS defects in the substrate the same way, see Eq.~\ref{eq:part_ratio}.
The difference in loss and frequency noise between the CPW and IDC can then be attributed to the effects of the inductor of the IDC resonator if one takes the stray capacitance in the inductor into account \cite{McRae2020}.

\begin{figure}[hbt]
\includegraphics{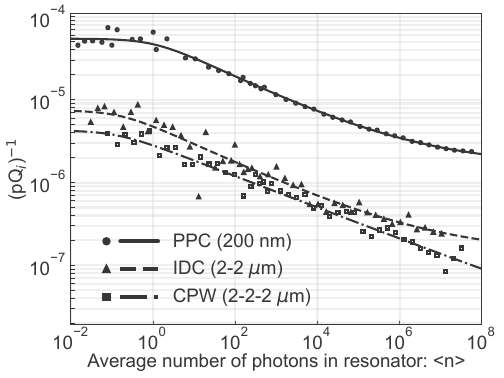}
  \caption{Measured loss tangent $\tan\delta_d=(pQ_i)^{-1}$, for a PPC-, IDC-, and CPW-based resonator. Scatter at high $Q_i$ is higher due to the uncertainty in fitting deep $S_{21}$ dips [$\text{min}(S_{21}) = Q/Q_i$]. We fit Eq.~\ref{eq:TLS_loss} to the measured data points with powers below the bifurcation point of the resonator.}
  \label{fig:3_loss}
\end{figure}

The microwave losses [$(pQ_i)^{-1}$] for one of each resonator type are plotted in Fig.~\ref{fig:3_loss}.
Since the layer thickness of the substrate TLS layer that gives the TLS-induced loss of the CPW an IDC \cite{Gao2008thesis, Barends2010} is unknown, the participation ratio, $p$, is unknown and we cannot extract $\tan\delta_d$ from the CPW and IDC measurements.
The CPW and IDC show similar noise levels, so the loss introduced by the inductor is negligible. 
The PPC on the other hand has roughly an order of magnitude higher $\tan\delta_{\rm{TLS}}$ than the CPW resonator, which shows that the PPC resonator is dominated by $\tan\delta_{\rm{TLS}}$ of the PPC and the contribution of the inductor is negligible.

\begin{figure}[hbt]
\includegraphics{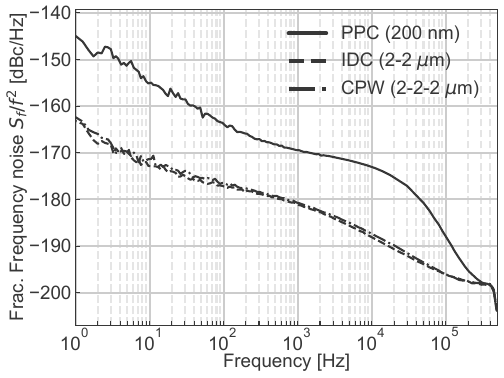}
  \caption{Resonator frequency noise, $S_f/f^2$, from the measured phase noise, $S_{\theta}$, power spectral density of each resonator at the same internal power, $P_{\rm{int}} = -50$ dBm. PPC has a plate area of 6000 $\mu\rm{m}^2$ (60$\times$100 \SI{}{\micro\meter\squared})}
  \label{fig:3_noise}
\end{figure}

The measured frequency noise spectra ($S_f/f^2$) for one of each resonator type are plotted in Fig.~\ref{fig:3_noise}.
The IDC and CPW resonators have near-identical noise spectra, while the PPC resonator has a 15--20-dB-higher noise level.
As before, we conclude that the contribution of the inductor is negligible, and the frequency noise of the PPC is dominated by the \textit{a}-$\mathrm{SiC}$:$\mathrm{H}$ PPC.
Note that the point at which the spectra change to $f^{-1/2}$ is at lower frequencies for the CPW and IDC resonators, around 10 Hz.
For the PPC resonators this is transition lies between $10^2$ and $10^3$ Hz, see Fig.~\ref{fig:3_noise}. 

\section{Frequency noise spectra analysis}
\label{sec:freq_noise_a}

\begin{figure}[hbt]
\includegraphics{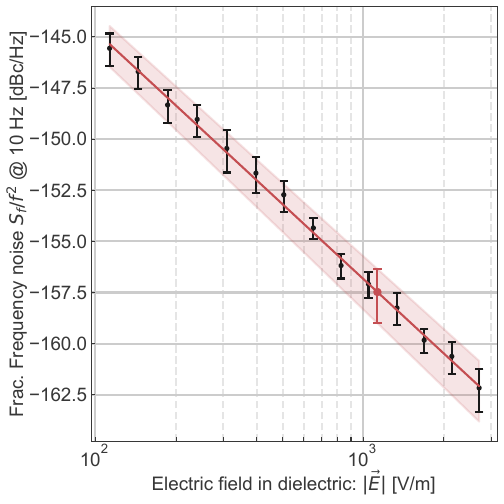}
  \caption{Fractional frequency noise at 10 Hz versus the internal electric field of the PPC. Each point is retrieved by fitting the $1/f$ region of the frequency noise spectra and calculating the 10-Hz point from the fitted parameter. Error bar of each point is the calculated uncertainty based on the standard-deviation error of the fit parameter. Red line is a fit with $a |\vec{E}|^{-b}$, with the uncertainty region based on the standard-deviation error of the fit. Red point is the retrieved frequency noise at 10 Hz and the desired electric field.}
  \label{fig:E_interp}
\end{figure}

Several analysis steps have to be taken to arrive at Fig.~\ref{fig:TLS_noise}(b) from the measured frequency noise spectra in Fig.~\ref{fig:TLS_noise}(a).
First, the $1/f$ region of the spectrum is fitted to a linear $af^{-b}$ relationship from 5 to 50 Hz.
The frequency noise at 10 Hz is then calculated with the fitted parameters $(a,b)$ and has an uncertainty given by the standard-deviation error on those parameters.
The result is the frequency noise at 10 Hz at the measured read powers for each KID.
The readout power is translated to the internal electric field inside the PPC through Eq.~\ref{eq:E-field}.
This yields the points in Fig.~\ref{fig:E_interp}, which follow the expected $|\vec{E}|^{-1}$ relationship \cite{Gao2008thesis, Faoro2015}.

To get the frequency noise at a desired electric field ($|\vec{E}_{des}|$), we fit the expected $a\cdot |\vec{E}|^{-b}$ relation to these data points, with $b\approx1$.
Using the fitted function parameters we obtain the frequency noise level at the desired field strength, with uncertainty given by the standard-deviation error in the fitted parameters ($a$,$b$).
The total error of the point $S_f(f = 10 \text{ Hz}, |\vec{E}| = |\vec{E}_{des}|)$ in Fig.~\ref{fig:E_interp} is the propagated error of the uncertainties in the $S_f(f)$ and $S_f(f = 10 \text{ Hz}, |\vec{E}|)$ fits.

\section{Temperature dependence}
\label{sec:Temp_dep}

The interacting tunneling model \cite{Faoro2015} predicts the following temperature-dependent TLS noise level.

\begin{equation}
	S_f \propto
	\begin{cases}
		T^{-(1+\mu)}, & T\ll \omega_r \text{ in weak field}\\
		T^{(1-\mu)/2}, & T\ll \omega_r \text{ in strong field}\\
		T^{\mu-1}, & T\gg \omega_r
	\end{cases}
	\label{eq:S_T}
\end{equation}

\begin{figure}[hbt]
  \includegraphics{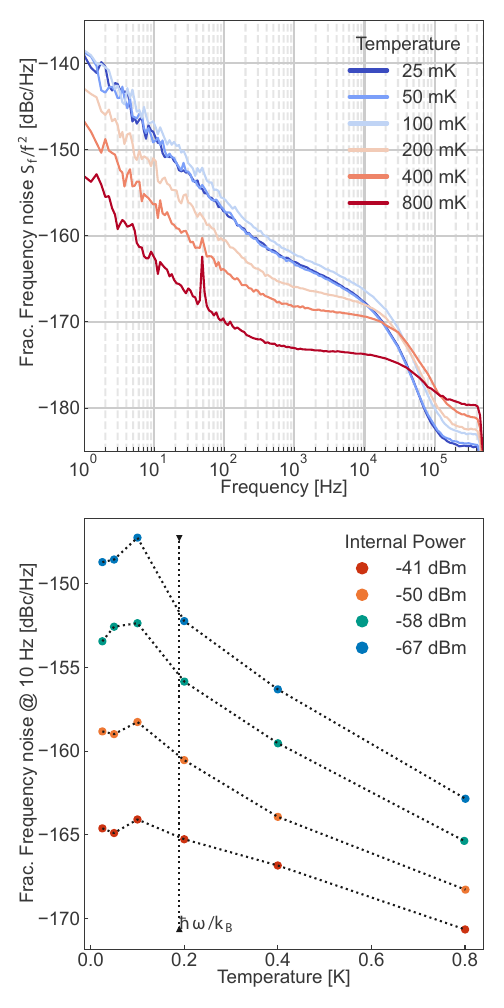}
  \caption{Fractional frequency noise of a PPC resonator for different bath temperatures and internal powers at 10 Hz. The internal powers fall within the strong field condition of Eq.~\ref{eq:S_T}. The vertical line is based on the resonance frequency at 100 mK: $T=\hbar\omega/k_{B}$. The dotted lines connect measured data points to highlight the backbend below $\hbar\omega=k_{B}T$.}
  \label{fig:Noise_T}
\end{figure}

We measure the frequency noise at temperatures between 25 mK and 1 K for all three resonator designs (PPC, IDC, CPW).
The frequency noise spectra at one internal power and the frequency noise at 10 Hz versus internal power for a PPC resonator are plotted in Fig.~\ref{fig:Noise_T}.
For all powers, the resonator is operated under the strong-field condition of Eq.~\ref{eq:S_T}, where one expects a smooth crossover between the limits of Eq.~\ref{eq:S_T} around $\hbar\omega=k_{B}T$.
In Fig.~\ref{fig:Noise_T}, the frequency noise indeed shows a backbend below $\hbar\omega=k_{B}T$.  
This temperature dependence is not explained by the STM \cite{Gao2008thesis} but can be explained by the interacting tunneling model presented in Ref.~\cite{Faoro2015}.

The spectrum at $f>$ 100 Hz follows an $f^{-1/2}$ relationship at the lowest temperatures. 
The slope of this region changes with temperature, resulting in an almost-white spectrum between 1 kHz and the resonator roll-off at 800 mK.
The $f>$ 100-Hz region roughly follows the temperature dependence of Eq.~\ref{eq:S_T}.
Neither the STM nor the interacting tunneling model explains the spectrum's $f^{-1/2}$ region nor its temperature dependence. 

\newpage
\bibliography{References}

\end{document}